\documentclass[aps,onecolumn]{revtex4}
\usepackage{epsf}
\usepackage{graphicx}
\usepackage{epstopdf}
\usepackage{tikz}
\usepackage{float}
\usepackage{amsmath}

\begin{document}
\title{Novel non-local effects in three-terminal hybrid devices with quantum dot}

\author{G. Micha\l ek$^{1}$,  T.\ Doma\'nski$^{2}$,
        B.R. Bu\l ka$^{1}$ and  K.I. Wysoki\'nski$^{2}$}
\affiliation{
       $^{1}$Institute of Molecular Physics, Polish Academy of Sciences, 
        ul. M. Smoluchowskiego 17, 60-179 Pozna\'n, Poland\\
       $^{2}$Institute of Physics, M. Curie-Sk\l odowska University, 
       pl. M. Curie-Sk{\l}odowskiej 1, 20-031 Lublin, Poland}

\date{\today}

\begin{abstract}
We predict strong non-local effects in the three-terminal hybrid device, comprising the quantum dot embedded between two conducting leads and third superconducting reservoir. They result from competition between the ballistic electron transfer and the crossed Andreev scattering. The non-local voltage induced in response to the 'driving' current changes the magnitude and sign upon varying the gate potential and/or coupling to the superconducting lead. Such effect is robust both in the linear and non-linear regimes, where the screening and the long-range interactions play significant role. This novel subgap transport is provided by the Shiba states and can be contrasted with much weaker non-local effects observed hitherto in the three-terminal 'planar' junctions.
\end{abstract}

\maketitle

\thispagestyle{empty}

Multi-terminal systems enable measurements of both the local and the nonlocal voltages/currents between selected electrode pairs~\cite{nazarov}. The non-local transport of charge~\cite{russo2005,cadden2006,brauer2010,webb2012,futturer2009,schindele2014}, heat~\cite{machon2013} and spin~\cite{noh2013} via hybrid devices consisting of the normal and superconducting reservoirs are currently of interest for the basic research and innovative applications. Electrons
traversing metal-superconductor interface are glued into the Cooper pairs, and conversely, the Cooper pairs are split into the individual electrons~\cite{rodero2011}. In both processes there emerge the entangled carriers, leading to nonlocal correlations. These effects can be amplified by inserting the quantum dots between the reservoirs~\cite{franceschi2010}. In this regard, the three-terminal structures are especially useful, because they allow for efficient splitting of the Cooper pairs~\cite{hofstetter2009,hermann2010,schindele2012}, give rise to spin filtering~\cite{braunecker2013}, generate the correlated spin currents~\cite{he2014}, separate the charge from heat currents~\cite{fazio2015}, enable realization of the exotic Weyl or Majorana-type quasi-particles~\cite{majorana}, etc.

Very spectacular non-local effects are provided by the crossed Andreev reflections (CAR), operating in a subgap regime. The 'driving' current applied to one side of the multi-terminal junction can yield either positive or negative nonlocal voltage response at the other interface, depending on a competition between the ballistic electron transfer (ET) and the CAR processes. Such changeover has been observed in three-terminal planar junctions~\cite{russo2005,cadden2006,brauer2010,webb2012}, using a piece of superconducting sample sandwiched between two conducting (normal or magnetic) electrodes. The induced non-local conductance, however, was much weaker from the local one in agreement with theoretical predictions~\cite{falci2001,golubev2007,melin2008}.

\begin{figure}[ht]
\centering
\includegraphics[width=0.7\linewidth]{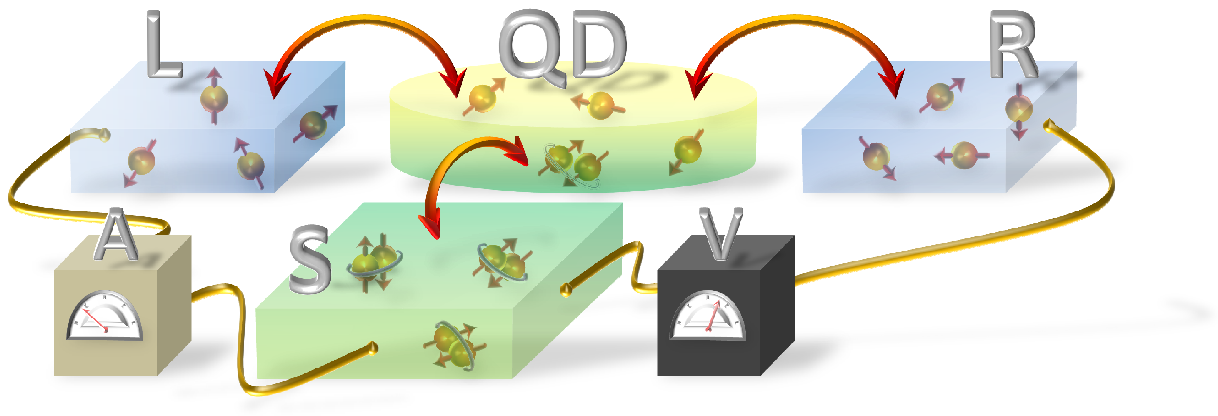}
\caption{(color online) Scheme of the three-terminal device consisting of two conducting leads ($L$ and $R$), superconducting reservoir ($S$) and the quantum dot ($QD$). The 'driving' current in the $L-QD-S$ loop induces the non-local voltage 'response' of the floating $R$ electrode.}
\label{fig1}
\end{figure}

Here we propose a different configuration, where the quantum dot is built into the three-terminal hybrid as sketched in Fig.~ \ref{fig1}. Proximity effect converts the quantum dot into, a kind of, superconducting grain and its subgap spectrum develops the, so called, Andreev or Shiba bound states~\cite{shiba}, which substantially enhance the non-local transport. We show that effective non-local conductance can be comparable to the local one and can change sign from the positive to negative values by increasing the coupling $\Gamma_S$ to superconducting electrode or by appropriate tuning of the gate potential. The gate potential is also controlling symmetry of this effect. Experimental tests of such effects should be feasible using the three-terminal architecture with such quantum dots as the carbon nano-tubes~\cite{pillet2013,schindele2014}, semiconducting nano-wires~\cite{lee2012,lee2014} or self-assembled InAs islands~\cite{deacon2010}.

\section*{Microscopic model}

Some aspects of the local and non-local transport properties for this three-terminal device could be inferred by extending the Landauer-B\"{u}ttiker approach~\cite{buttiker1988,lambert1993,allsopp1994,lambert1998,morten2006} (see the 1-st subsection of Methods). On a microscopic level, we describe this system in the tunneling approximation~\cite{mahan2000} by the Hamiltonian
\begin{eqnarray}
H &=& \sum_{\alpha = L, R} \sum_{k, \sigma} \epsilon_{\alpha k} c_{\alpha k \sigma}^\dag c_{\alpha k \sigma} + \sum_\sigma \left[ \epsilon_0 - e U(\mathbf{r}) \right] d_\sigma^\dag d_\sigma + \sum_{\alpha, k, \sigma} \left( t_\alpha c_{\alpha k \sigma}^\dag d_\sigma + t_\alpha^* d_\sigma^\dag c_{\alpha k \sigma} \right) \nonumber \\
&+& \sum_{k, \sigma} \epsilon_{S k} c_{S k \sigma}^\dag c_{S k \sigma} - \sum_k \Delta \left( c_{S -k \uparrow}^\dag c_{S k \downarrow}^\dag + c_{S k \downarrow} c_{S -k \uparrow} \right)
\label{eq-Ham}
\end{eqnarray}
with standard notation for the annihilation (creation) operators of the itinerant $c_{\alpha k \sigma}^{(\dagger)}$ and localized dot $d_\sigma^{(\dagger)}$ electrons. The first term describes the left ($\alpha = L$) and the right ($\alpha = R$) conducting leads. The subsequent term refers to the quantum dot ($QD$) with its energy level $\epsilon_0$ shifted by the long-range potential $U(\mathbf{r})$. Hybridization between the $QD$ and itinerant electrons is characterized by the matrix elements $t_{\alpha}$. The last two terms in (\ref{eq-Ham}) correspond to the BCS-type superconducting reservoir with an isotropic energy gap $\Delta$. Addressing here the subgap (low-energy) transport we assume the constant tunneling rates $\Gamma_\alpha = 2 \pi \sum_k |t_\alpha|^2 \delta (E - \epsilon_{\alpha k}) = 2 \pi |t_\alpha|^2 \rho_\alpha$, where $\rho_\alpha$ is the (normal state) density of states of $\alpha$ lead. In what follows, we assume the superconducting gap $\Delta$ to be the largest energy scale in the problem.

\section*{Subgap charge transport}

The charge current $J_\alpha$ flowing from an arbitrary lead $\alpha = \{ L, R, S \}$ can be evaluated using the Heisenberg equation $J_\alpha \equiv e \langle \dot{N}_\alpha \rangle = - i e / \hbar \langle [N_\alpha, H] \rangle$~\cite{EOM}.
In particular, the current $J_{L (R)}$ from the normal $L$ ($R$) electrode is given by~\cite{EOM}
\begin{equation}
\label{eq-curog}
J_{L (R)} = \frac{4e}{h} \int d E \Gamma_{L (R)} \Im \left[ f_{L (R)} G_{11}^r + \frac{1}{2} G_{11}^< \right] \; ,
\end{equation}
where $G_{11}^r$ and $G_{11}^<$ denote the matrix elements (in the Nambu representation) of the retarded and lesser $QD$ Green functions, respectively. From now onwards we consider the current $J_L$ focusing on the subgap voltage, smaller than the energy gap $|\Delta|$. In such regime there are possible: the ballistic electron transfer (ET) from $L$ to $R$ electrode, the direct Andreev reflection (DAR) when electron from $L$ lead is converted into the Cooper pair in $S$ reservoir and hole is scattered to $L$ electrode, and the crossed Andreev reflection (CAR) which is similar to DAR except that hole is scattered to $R$ electrode. They can be expressed as~\cite{michalek2013}
\begin{eqnarray}
\label{curr-gen-ET}
J_L^{ET} &=& \frac{2e}{h} \int d E \; \Gamma_L \Gamma_R \; |G_{11}^r|^2 \left( f_L - f_R \right) , \\
\label{curr-gen-DAR}
J_L^{DAR} &=& \frac{2e}{h} \int d E \; \Gamma_L^2 |G_{12}^r|^2 \left( f_L - \tilde{f}_L \right) , \\
\label{curr-gen-CAR}
J_L^{CAR} &=& \frac{2e}{h} \int d E \; \Gamma_L \Gamma_R |G_{12}^r|^2 \left( f_L - \tilde{f}_R \right) ,
\end{eqnarray}
where $f_\alpha \equiv f_\alpha (E) = \{ \exp [ (E - \mu_\alpha) / k_B T ] + 1 \}^{-1}$ and $\tilde{f}_\alpha \equiv \tilde{f}_\alpha (E) = 1 - f_\alpha (-E) = \{ \exp [ (E + \mu_\alpha) / k_B T ] + 1 \}^{-1}$ are the Fermi-Dirac distribution functions for electrons and holes, respectively. Let us remark, that only the ET (\ref{curr-gen-ET}) and CAR (\ref{curr-gen-CAR}) contributions lead to the non-local effects, because they depend on the chemical potentials of both conducting ($L$ and $R$) electrodes. Since these ET and CAR processes deliver different types of the charge carriers to the right electrode, the induced voltage $V_R$ would be a probe of the dominant transport mechanism.

\begin{figure}[ht]
\centering
\includegraphics[width=0.47\linewidth]{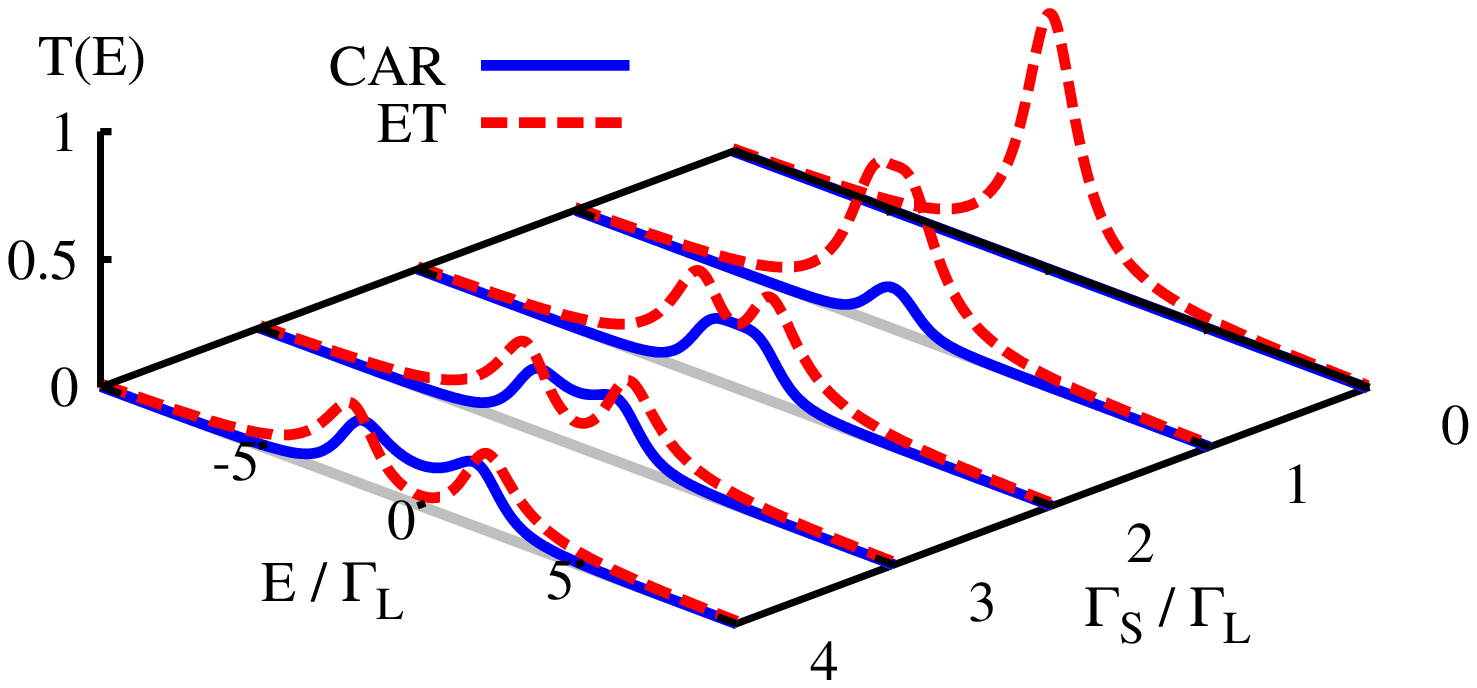} \hspace{0.5cm}
\includegraphics[width=0.47\linewidth]{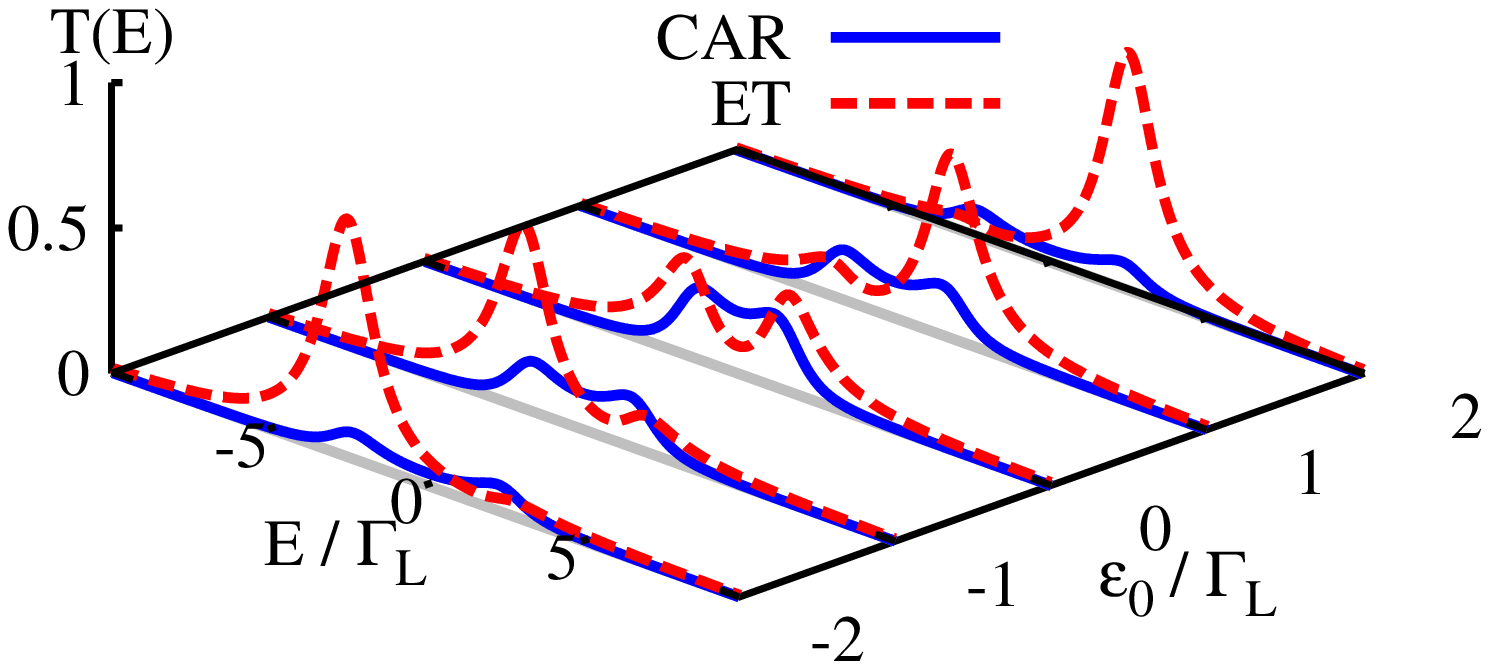}
\caption{(color online) Transmissions of the ET (dashed lines) and CAR (solid lines) transport channels obtained at zero temperature for $\Gamma_R = \Gamma_L$. The left panel refers to $\epsilon_0 = 0$ and the right one to $\Gamma_S = 3 \Gamma_L$.}
\label{Transmissions}
\end{figure}

Relationship between the ET and CAR processes can be inspected by studying their transmissions, defined as $T^{ET} (E) = \Gamma_L \Gamma_R |G_{11}^r (E)|^2$ and $T^{CAR} (E) = \Gamma_L \Gamma_R |G_{12}^r (E)|^2$ (see Fig.~\ref{Transmissions}).
Deep in a subgap regime (i.e.\ for $|E| \ll |\Delta|$) the Green function $\hat{G}^r (E)$ simplifies to the familiar BCS structure \cite{yamada2011}. 
Its diagonal part is given by $G_{11}^r (E) = u^2 / \left[ E - E_A + \frac{i}{2} \Gamma_N \right] + v^2 / \left[ E + E_A + \frac{i}{2} \Gamma_N \right]$ with the quasi-particle energy $E_A = \sqrt{\epsilon_0^2 + \Delta_{QD}^2}$, where $\Delta_{QD} = \frac{1}{2} \Gamma_S$. Subgap spectrum consists thus of two Shiba states at $\pm E_A$ whose spectral weights are $u^2 = \frac{1}{2} \left[ 1 + \Delta_{QD} / E_A \right]$ and $v^2 = 1 - u^2$ with the quasiparticle broadening $\Gamma_N = \Gamma_L + \Gamma_R$. The single electron transmission $T^{ET} (E)$ is a quantitative measure of this subgap spectrum.
The left panel in Fig.~\ref{Transmissions} illustrates evolution of the Shiba states upon increasing the coupling $\Gamma_S$ while the right panel shows a transfer of the corresponding spectral weights $u^2 \leftrightarrow v^2$ upon changing the $QD$ level $\epsilon_0$ by an applied gate voltage.

Transmission of the anomalous CAR channel, on the other hand, depends on the off-diagonal part of the matrix Green function $G_{12}^r (E) = uv / \left[ E - E_A + \frac{i}{2} \Gamma_{N} \right] - uv / \left[ E + E_A + \frac{i}{2} \Gamma_{N} \right]$, where $uv = \frac{1}{2} \Delta_{QD} / E_A$. It also has maxima around the same Shiba states $\pm E_A$ but with a different amplitude, sensitive to the induced pairing $\langle d_\downarrow d_\uparrow \rangle$. This is a reason why $T^{CAR} (E)$ quickly diminishes whenever $\Gamma_S$ is decreased or the $QD$ level $\epsilon_0$ departs from $\mu_S = 0$ (solid lines in Fig.~\ref{Transmissions}).

Confronting both these transmissions reveals that the non-local transport predominantly comes from the CAR process when the coupling $\Gamma_S$ (to superconducting electrode) is sufficiently strong and the $QD$ level $\epsilon_0$ is close to the chemical potential $\mu_S$. Otherwise, the non-local effects are dominated by the single electron tunneling (ET). The related changeover can be detected by measuring the voltage $V_R$ in the floating $R$ electrode, in response to the current in the $L-QD-S$ branch. Such voltage $V_R$ can vary between the positive and negative values and the non-local resistance can be tuned by the gate potential lifting/lowering the Shiba energies.

\section*{Linear response}

Practical realizations of the setup (Fig.~\ref{fig1}) would allow to measure the local and the non-local resistances/conductances within the four-probe scheme~\cite{buttiker1988,lambert1993,allsopp1994,lambert1998}, where the potentials and currents are treated on equal footing (see the Method). In a weak perturbation limit the response would be linear
\begin{equation}
J_L \simeq e \mathcal{L}_{LR}^{ET} (V_L - V_R) + e \mathcal{L}_{LL}^{DAR} [(V_L - V_S) - (V_S - V_L)] + e \mathcal{L}_{LR}^{CAR} [(V_L - V_S) - (V_S - V_R)] \; .
\end{equation}
%
\begin{figure}[ht]
\centering
\includegraphics[width=0.47\linewidth]{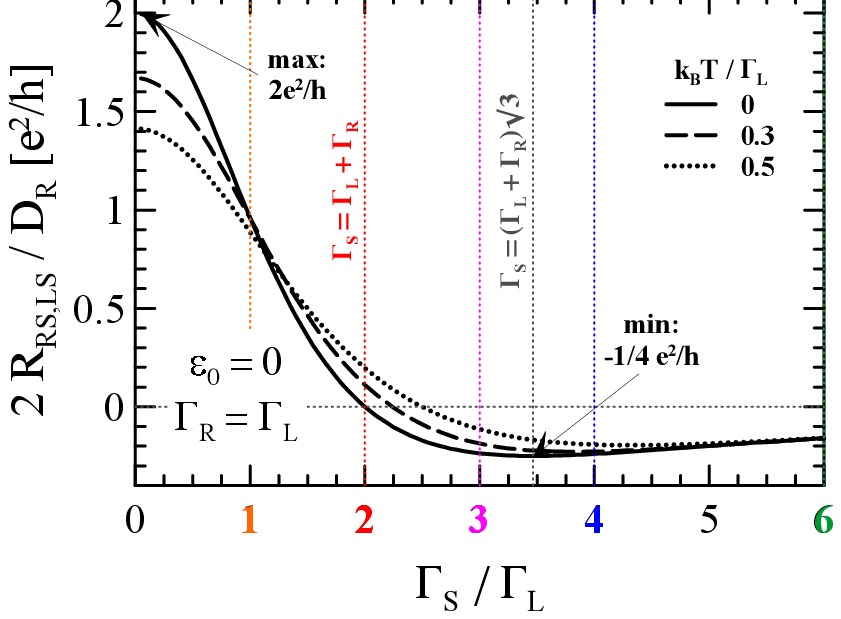} \hspace{0.5cm}
\includegraphics[width=0.47\linewidth]{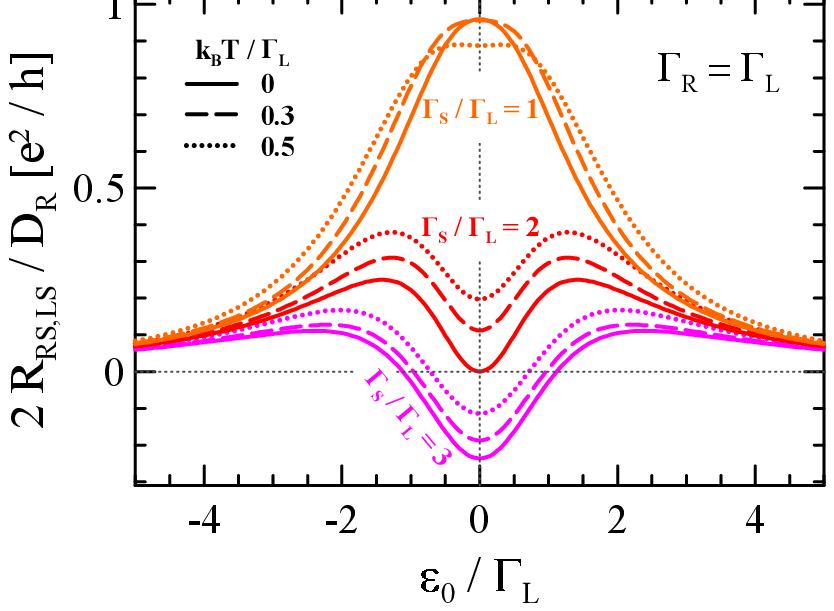}
\caption{(color online) The non-local resistance $2 R_{RS, LS} / D_R$ as a function of $\Gamma_S / \Gamma_L$ ratio (left panel) and the $QD$ dot level $\epsilon_0$ (right panel) obtained in the linear limit for three representative temperatures.}
\label{res_linear}
\end{figure}
The coefficients $\mathcal{L}_{ij}^\beta$ for $\beta = $ ET, DAR or CAR can be determined from the equations (\ref{curr-gen-ET}-\ref{curr-gen-CAR}) and they read
\begin{equation}
\mathcal{L}_{ij}^\beta = \frac{2e}{h} \int dE \; T^\beta (E) \left[ - \frac{\partial f}{\partial E} \right] \; .
\end{equation}
%
At zero temperature $- \frac{\partial f}{\partial E} \approx \delta (E)$, hence $\mathcal{L}_{ij}^\beta$ depend on the transmissions $T^\beta (E \rightarrow 0)$.

Treating the potential $V_S$ as a reference level we analyze the induced voltage $V_R$ in response to the 'driving' current $J_L \equiv J_{LS}$. The local resistance $(V_L - V_S) / J_{LS} = R_{LS, LS}$ is due to the DAR processes whereas the non-local one $(V_R - V_S) / J_{LS} = R_{RS, LS}$ results from the single electron tunneling (ET) competing with the anomalous crossed Andreev reflection (CAR) processes. Fig.~\ref{res_linear} shows this non-local resistance $R_{RS, LS}$ normalized with respect to $D_R / 2 = R_{LS, LR} R_{RS, LS} + R_{LS, LR} R_{RS, RL} + R_{RS, LS} R_{RS, RL}$ [defined by equation (\ref{denominator}) in Methods].
The left panel shows that $R_{RS, LS}$ has a negative sign (signifying the dominant CAR processes) only for sufficiently strong coupling $\Gamma_S > \Gamma_N$. This is a straightforward consequence of the (zero-energy) ET and CAR transmissions (Fig.~\ref{Transmissions}). The right panel of Fig.~\ref{res_linear} displays the non-local resistance versus the $QD$ level $\epsilon_0$. In the linear regime the negative nonlocal resistance occurs when $\epsilon_0 \sim \mu_S$ for sufficiently strong coupling $\Gamma_S > \Gamma_N$. Since $\Gamma_S$ and $\epsilon_0$ can be experimentally varied in the realizations of the superconducting-metallic devices with the quantum dots~\cite{pillet2013,schindele2014,lee2012,lee2014,deacon2010}, such qualitative changes should be observable.

\subsection*{Beyond the linear response limit}

To confront these findings with the non-local effects observed so far in the 'planar' junctions~\cite{russo2005,cadden2006,brauer2010,webb2012} we now go beyond the linear response framework. For arbitrary value of the 'driving' voltage $V_L$ we computed self-consistently $V_R$, guaranteeing the net current $J_R$ to vanish. Under such non-equilibrium conditions the long-range potential $U (\mathbf{r})$ plays an important role in the transport when the charges pile up in the electrodes and the quantum dot~\cite{altshuler1985}. It affects the chemical potentials and the injectivities of the leads and contributes to the screening effect~\cite{buttiker1993,buttiker1997,ma1998,wang2001}. The potential $U (\mathbf{r})$ has to be properly adjusted, depending on specific polarization of the system~\cite{buttiker1997} (for details see the 2-nd subsection of Methods).

\begin{figure}[ht]
\centering
\includegraphics[width=0.47\linewidth]{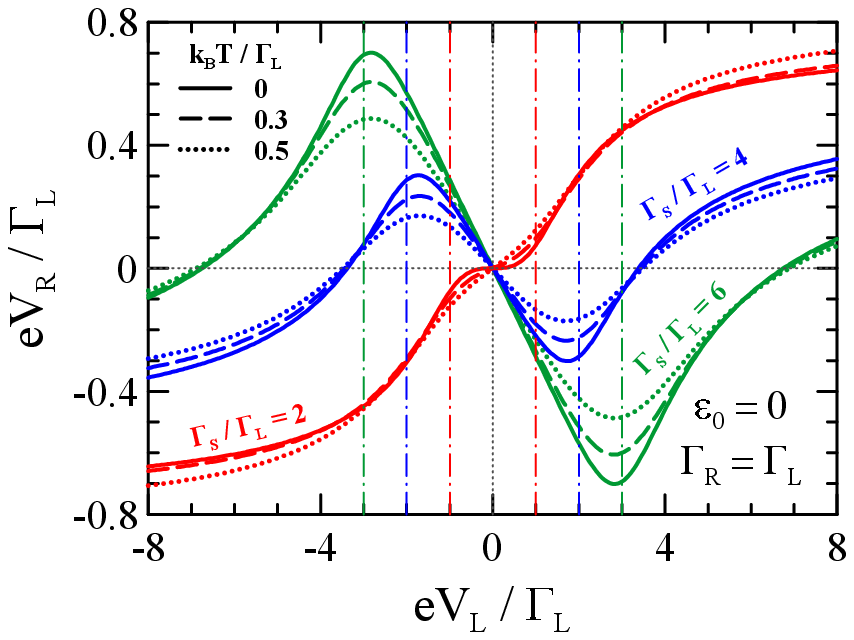} \hspace{0.5cm}
\includegraphics[width=0.47\linewidth]{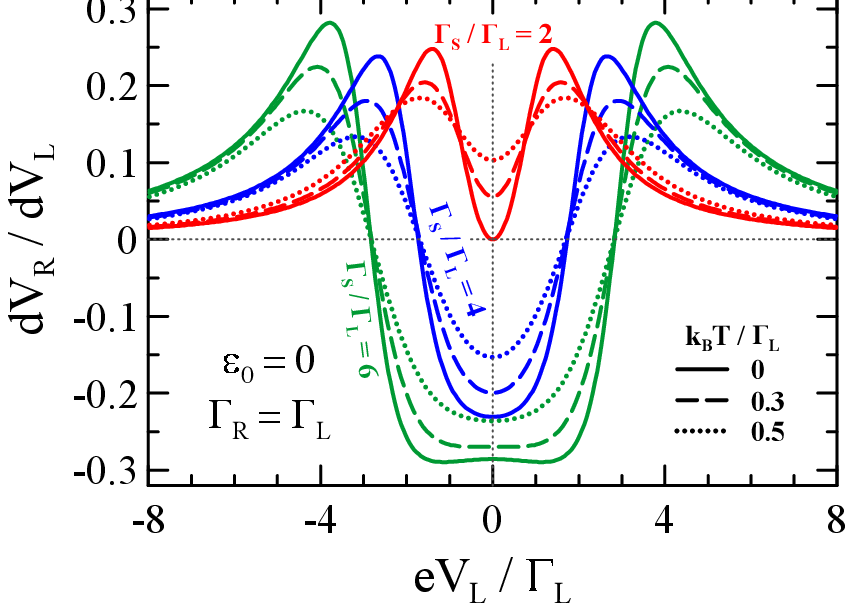}
\caption{(color online) The non-local voltage $V_R$ (left panel) and its derivative $d V_R / d V_L$ (right panel) induced in the floating $R$ lead in response to the 'driving' voltage $V_L$.}
\label{vr-vs-vl}
\end{figure}

Figure~\ref{vr-vs-vl} shows the induced non-local voltage $V_R$ and its derivative with respect to $V_L$ for several couplings $\Gamma_S$ and temperatures, obtained for $U (\mathbf{r}) = 0$. At low voltage $|V_L|$ the induced potential $V_R$ is proportional to $V_L$, as we discussed in the linear response regime (Fig.~\ref{res_linear}). Upon increasing the 'driving' voltage $|V_L|$ the Shiba states $\pm E_A$ (indicated by vertical lines in Fig.~\ref{vr-vs-vl}) are gradually activated, amplifying the non-local processes. For $\Gamma_S > \Gamma_N$ we hence observe local minima (maxima) of $V_R$ at the quasiparticle energies $E_A$ ($- E_A$). Further increase of $|V_L|$ leads to revival of the dominant ET channel. The derivative $d V_R / d V_L$, which is related to the ratio of the local and non-local differential resistances $R_{LS, RS} / R_{LS, LS}$, can be measured by the standard lock-in method. Our results  differ qualitatively from the properties of the planar junctions (where the ET and CAR dominated regions are completely interchanged)~\cite{russo2005,cadden2006,brauer2010,webb2012} because the non-local transport  occurs through the Andreev states, that are localized at two normal-superconductor interfaces 
separated by a distance $d$ comparable to the coherence length of superconductor. 
In consequence, the anomalous CAR transport is possible only for $eV_{L}$ exceeding
the characteristic Thouless energy~\cite{falci2001,golubev2007,melin2008}. 

\begin{figure}[ht]
\centering
\includegraphics[width=0.47\linewidth]{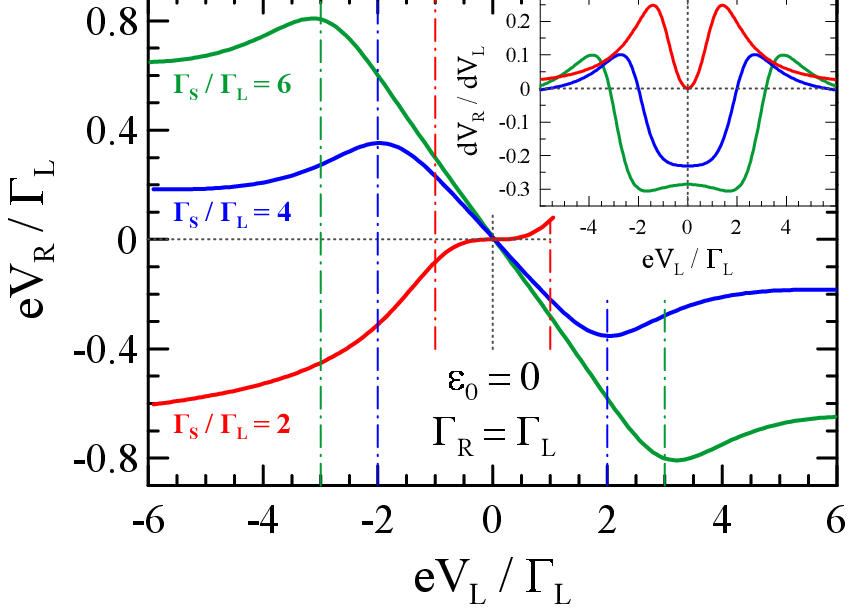} \hspace{0.5cm}
\includegraphics[width=0.47\linewidth]{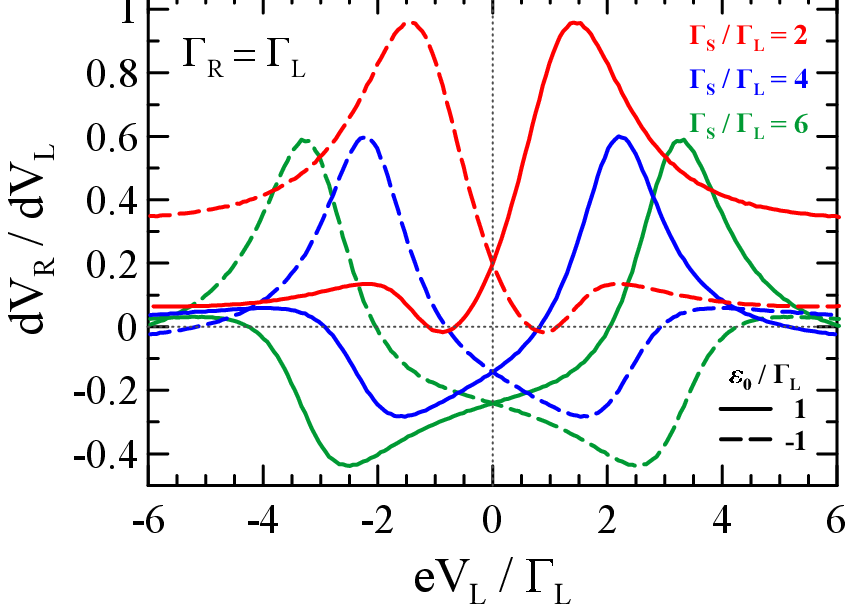}
\caption{(color online) The non-local voltage $V_R$ and its derivative with respect to $V_L$ obtained at low temperature for $\epsilon_0 = 0$ (left panel) taking into account the screening effects $U(\textbf{r})$. The lower panel shows $d V_R / d V_L$ for $\epsilon_0 / \Gamma_L = \pm 1$.}
\label{nl-vr-vs-vl}
\end{figure}

Feedback effect of the long-range potential $U (\mathbf{r}) = U_{eq} + \sum_\alpha u_\alpha V_\alpha$ (where $U_{eq}$ denotes the equilibrium value incorporated into $\epsilon_0$) is illustrated in Fig.~\ref{nl-vr-vs-vl}. The quantitative changes are observed for all voltages, however, the qualitative behaviour is similar to that found in the linear regime (Fig.~\ref{vr-vs-vl}).
The screening effects and injectivities are calculated here in the self-consistent way~\cite{altshuler1985,buttiker1993,buttiker1997,ma1998,wang2001} (discussed in the 2-nd subsection of Methods). This selfconsistent treatment of $U (\textbf{r})$ partly suppresses both the non-local voltage $V_R$ and $d V_R / d V_L$. The right panel of Fig.~\ref{nl-vr-vs-vl} shows $d V_R / d V_L$ with respect to $V_L$ outside the particle-hole symmetry point, i.e. for $\epsilon_0 = \pm \Gamma_L$. These asymmetric curves can be practically obtained by applying the gate potential to the quantum dot.

\section*{Summary and Outlook}

We proposed the three-terminal hybrid device, where the quantum dot is tunnel-coupled to two normal and another superconducting electrode, for implementation of the efficient non-local transport properties. We investigated such effects in the linear and non-linear regimes. We found that in the both cases the non-local resistance/conductance can change from the positive (dominated by the usual electron transfer) to negative values (dominated by the crossed Andreev reflections) upon varying the coupling to superconducting electrode $\Gamma_S$ and tuning
the $QD$ level $\epsilon_0$.

This nano-device would enable realization of the strong non-local conductance
(comparable to the local one) by activating the Shiba states formed at sub-gap
energies $\pm E_A$. They substantially enhance all the transport channels, in
particular promoting the CAR mechanism (manifested by the negative non-local
conductance/resistance) when the coupling to superconducting electrode is
strong $\Gamma_S > \Gamma_L + \Gamma_R$.  We predict the negative non-local
conductance/resistance both, in the linear regime and beyond it. For the latter
case such behavior would be observable exclusively in the low bias voltage
regime $|V_{L}| < E_{A}/e$ capturing the Shiba states. The quantum dot level
$\epsilon_0$ (tunable by the gate potential) can additionally control asymmetry
of the non-linear transport properties, affecting the CAR transmission $T^{CAR}
(\pm E_A) \propto \left[ 1 + \left( 2 \epsilon_0 / \Gamma_S \right)^2 \right]^{-1}$.

Strong non-local properties of the nano-device (shown in figure 1) can be contrasted
with the previous experimental measurements for the three-terminal planar junctions
(consisting of two $N-S$ interfaces separated by a superconducting mesoscopic
island)  \cite{russo2005,cadden2006,brauer2010,webb2012}.
Russo \textit{et al.}~\cite{russo2005} reported evolution from the positive to
negative non-local voltage $V_R$ induced in response to the 'driving' bias $V_L$.
At low $V_L$ the ET processes dominated, whereas for higher $V_L$ the CAR took over.
The sign change of $V_R$ occurred at voltage $V_L$ related to the Thouless energy
(such changeover completely disappeared when a width of the tunneling region
via the superconducting sample exceeded the coherence length). Similar weak
negative non-local resistance/conductance has been observed in the spin valve
configurations \cite{brauer2010,webb2012}. In the planar junctions
the non-local conductance was roughly 2 orders of the magnitude weaker
than the local one~\cite{brauer2010}.

Summarizing, we proposed the nanoscopic three-terminal device for the tunable
(controllable) and very efficient non-local conductance/resistance ranging between
the positive to negative values. Our theoretical predictions can be verified
experimentally (in the linear response regime and beyond it) using any
quantum dots \cite{pillet2013,schindele2014,lee2012,lee2014,deacon2010}
attached between one superconducting and two metallic reservoirs.
Such measurements are called for.

\section*{Appendix A: Landauer-B\"{u}ttiker formalism}


The four-point method~\cite{buttiker1988} is well established technique for measuring the resistance in a ballistic regime. Voltage $V_{kl}$ measured between $k$ and $l$ electrodes in response to the current $J_{ij}$ between $i$ and $j$ electrodes defines the local ($ij = kl$) or non-local ($ij \neq kl$) resistance via
\begin{equation}
R_{ij, kl} \equiv \frac{V_k - V_l}{J_{ij}} = \frac{\mu_k - \mu_l}{e J_{ij}} = \frac{\Delta \mu_{kl}}{e J_{ij}} \; ,
\label{eq:def-opory}
\end{equation}
where $\Delta \mu_{kl} = \mu_k - \mu_l$ is a difference between the chemical potentials of $k$ and $l$ electrodes. The formalism has been later extended by Lambert \textit{et al.}~\cite{lambert1993,allsopp1994} to systems, where electron tunneling occurs between one or more superconductors. The current from $i$-th lead depends on the chemical potential $\mu_S$ of superconducting reservoir, because the scattering region acts as a source or sink of quasi-particle charge due to the Andreev reflection (see e.g. Ref.~[\onlinecite{lambert1998}]).

Adopting this approach, we analyze here the local and non-local transport properties of the three-terminal hybrid system consisting of two normal ($L$ and $R$) leads coupled through the quantum dot with another superconducting ($S$) electrode. We consider the charge transport driven by small (subgap) voltages $eV_{kl} \equiv \Delta \mu_{kl} = \mu_k - \mu_l \ll \Delta$, when the single electron transfer to the superconductor is prohibited. In this limit the net current flowing from the normal $L$ electrode consists of the following three contributions
\begin{eqnarray}
J_L = \mathcal{L}_{LR}^{ET} \left( \mu_L - \mu_R \right) + \mathcal{L}_{LL}^{DAR} \left[ (\mu_L - \mu_S) - (\mu_S - \mu_L) \right] + \mathcal{L}_{LR}^{CAR} \left[ (\mu_L - \mu_S) - (\mu_S - \mu_R) \right] \; .
\label{eq:jsc}
\end{eqnarray}
The linear coefficient $\mathcal{L}_{LR}^{ET}$ refers to the processes transferring single electrons between metallic $L$ and $R$ leads. We call this process as the electron transfer (ET). The other term with $\mathcal{L}_{LL}^{DAR}$ corresponds to the direct Andreev reflection, when electron from the normal $L$ lead is converted into the Cooper pair (in $S$ electrode) reflecting a hole back to the same lead $L$. The last coefficient $\mathcal{L}_{LR}^{CAR}$ describes the non-local crossed Andreev reflection, involving all three electrodes when a hole is reflected to the second $R$ lead. In the subgap regime the competing ET and CAR channels are responsible for the non-local transport properties.

In the same way as (\ref{eq:jsc}) one can express the current $J_R$. By symmetry reasons we have $\mathcal{L}_{RL}^{ET} = \mathcal{L}_{LR}^{ET}$ and $\mathcal{L}_{RL}^{CAR} = \mathcal{L}_{LR}^{CAR}$, whereas the charge conservation (Kirchoff's law) implies $J_S = - J_L - J_R$. From these linear response expressions  one can estimate the relevant local and non-local resistances (\ref{eq:def-opory}), assuming arbitrary configurations of the applied currents and induced voltages. Experimental measurements of such resistances (\ref{eq:def-opory}) can be done, treating one of the electrodes as a voltage probe. In our three-terminal device with the quantum dot we can assume either the metallic or superconducting electrode to be floating. We now briefly discuss both such options.

\subsubsection*{Floating metallic electrode}

We assume that the superconducting lead $S$ is grounded and treat the metallic electrode (say $L$) as a voltage probe. This means that the net current vanishes $J_L = 0$ and, from the charge conservation, one finds $J_R = -J_S \equiv J_{RS}$. In the linear response regime \eqref{eq:jsc} implies the following potential differences
\begin{eqnarray}
\frac{\Delta \mu_{RL}^L}{e J_{RS}} & \equiv & R_{RS, RL} = \frac{\mathcal{L}_{LL}^{DAR} + \mathcal{L}_{LR}^{CAR}}{e D} \; ,
\label{eq:s1} \\
\frac{\Delta \mu_{LS}^L}{e J_{RS}} & \equiv & R_{RS, LS} = \frac{\mathcal{L}_{LR}^{ET} - \mathcal{L}_{LR}^{CAR}}{2 e D} \; ,
\label{eq:s2} \\
\frac{\Delta \mu_{RS}^L}{e J_{RS}} & \equiv & R_{RS, RS} = \frac{\mathcal{L}_{LR}^{ET} + 2 \mathcal{L}_{LL}^{DAR} + \mathcal{L}_{LR}^{CAR}}{2 e D} = \frac{\Delta \mu_{LS}^L} {e J_{RS}} + \frac{\Delta \mu_{RL}^L}{e J_{RS}} = R_{RS, RL} + R_{RS, LS} \; ,
\label{eq:s3}
\end{eqnarray}
with a common denominator
\begin{eqnarray}
D = \mathcal{L}_{LR}^{ET} ( \mathcal{L}_{LL}^{DAR} + 2 \mathcal{L}_{LR}^{CAR} + \mathcal{L}_{RR}^{DAR} ) + \mathcal{L}_{LR}^{CAR} ( \mathcal{L}_{LL}^{DAR} + \mathcal{L}_{RR}^{DAR}) + 2 \mathcal{L}_{LL}^{DAR} \mathcal{L}_{RR}^{DAR} \; .
\label{denominator}
\end{eqnarray}
According to the definition (\ref{eq:def-opory}) and using (\ref{eq:s1}-\ref{eq:s3}) we obtain the local ($R_{RS, RS}$) and non-local ($R_{RS, RL}$, $R_{RS, LS}$) resistances for the floating $L$ lead. Let us notice, that a sign of the non-local resistance $R_{RS, LS}$ depends on a competition between the normal electron transfer (ET) and the crossed Andreev reflections (CAR). The local resistance $R_{RS, RS}$ is in turn a sum of the non-local resistances $R_{RS, RL}$ and $R_{RS, LS}$. For the configuration, where the other ($R$) metallic lead is floating we obtain the equations similar to (\ref{eq:s1}-\ref{eq:s3}) with the exchanged indices $L \leftrightarrow R$.

\subsubsection*{Floating superconducting electrode}

We encounter a bit different situation, assuming the superconducting $S$ electrode to be floating (i.e. $J_S = 0$). The charge conservation $J_L = - J_R \equiv J_{LR}$ and Eq.~\eqref{eq:jsc} imply
\begin{eqnarray}
\label{eq:s4}
\frac{\Delta \mu_{LS}^S}{e J_{LR}} & \equiv & R_{LR, LS} = \frac{L_{RR}^{DAR} + L_{LR}^{CAR}}{e D} = R_{LS, LR} \; , \\
\label{eq:s5}
\frac{\Delta \mu_{SR}^S}{e J_{LR}} & \equiv & R_{LR, SR} = \frac{L_{LL}^{DAR} + L_{LR}^{CAR}}{e D} = R_{RS, RL} \; , \\
\label{eq:s6}
\frac{\Delta \mu_{LR}^S}{e J_{LR}} & \equiv & R_{LR, LR} = \frac{L_{LL}^{DAR} + 2 L_{LR}^{CAR} + L_{RR}^{DAR}}{e D} = \frac{\Delta \mu_{LS}^S}{e J_{LR}} + \frac{\Delta \mu_{SR}^S} {e J_{LR}} = R_{LR, LS} + R_{LR, SR} = R_{LS, LR} + R_{RS, RL} \; .
\end{eqnarray}
We notice some analogy between the resistances (\ref{eq:s4}-\ref{eq:s6}) and the previous expressions (\ref{eq:s1}-\ref{eq:s3}). The significant difference appears between the non-local resistances $R_{RS, LS}$ (\ref{eq:s2}) and $R_{LR, SR}$ (\ref{eq:s5}). Because of a minus sign in (\ref{eq:s2}) the former configuration seems to be more sensitive for probing the local versus non-local transport properties.

\subsubsection*{Remarks on the determination of partial conductances}

Measurements of the local/non-local resistances provide information about the competition between various tunneling processes. Similar information can be also deduced about the linear coefficients $\mathcal{L}_{ij}^\beta$. Let's combine the results obtained for $L$ (or $R$) and $S$ floating electrodes. We have three independent equations, but we have to determine four coefficients
\begin{eqnarray}
\mathcal{L}_{LL}^{DAR} + \mathcal{L}_{LR}^{ET} &=& \frac{R_{LR, RS} - 2 R_{RS, LS}}{e D_R} \; , \nonumber \\
\mathcal{L}_{RR}^{DAR} + \mathcal{L}_{LR}^{ET} &=& - \frac{R_{LR, LS} + 2 R_{RS, LS}}{e D_R} \; , \\
\mathcal{L}_{LR}^{CAR} - \mathcal{L}_{LR}^{ET} &=& \frac{2 R_{RS, LS}} {e D_R} \; . \nonumber
\end{eqnarray}
In general, we thus cannot obtain a complete information about all conductances from the separate measurements of the currents and voltages. This situation differs from the case when the quantum dot is coupled to all three normal electrodes, where electrical transport can be characterized only by three conductances.

Fortunately, for the case with asymmetric couplings $\Gamma_R \neq \Gamma_L$ the measurements can unambiguously determine the partial conductances
\begin{eqnarray}
\mathcal{G}_{LR}^{ET} \equiv e \mathcal{L}_{LR}^{ET} & = & - \frac{\Gamma_L^2 ( R_{LR, LS} + 2 R_{RS, LS} ) + \Gamma_R^2 ( R_{LR, RS} - 2 R_{RS, LS} )}{( \Gamma_L^2 - \Gamma_R^2 ) D_R} \; , \\
\mathcal{G}_{LL}^{DAR} \equiv e \mathcal{L}_{LL}^{DAR} &=& \frac{\Gamma_L^2 ( R_{LR, LS} + R_{LR, RS} )} {( \Gamma_L^2 - \Gamma_R^2 ) D_R} \; , \\
\mathcal{G}_{RR}^{DAR} \equiv e \mathcal{L}_{RR}^{DAR} &=& \frac{\Gamma_R^2}{\Gamma_L^2} \mathcal{G}_{LL}^{DAR} \; , \\
\mathcal{G}_{LR}^{CAR} \equiv e \mathcal{L}_{LR}^{CAR} &=& - \frac{\Gamma_L^2 R_{LR, LS} + \Gamma_R^2 R_{LR, RS}}{( \Gamma_L^2 - \Gamma_R^2 ) D_R} \; .
\end{eqnarray}
Some inconvenience is related to the fact the tunneling rates $\Gamma_L$, $\Gamma_R$ must be measured as well.

\subsection*{Appendix B: Non-linear transport}

The non-linear effects are of vital importance in the transport studies of nanostructures \textit{inter alia} due to limited screening of charge and access to far from equilibrium states of the system. Non-equilibrium transport driven by the voltage $V_L$ (beyond the linear regime) in nanostructures is accompanied by substantial redistribution of the charges. This affects the occupancy of the quantum dot and leads to piling up of the charge in the electrodes. By long range Coulomb interactions the charge redistributions backreact on the transport properties. We shall address this effect in some detail.

Let's note that we are considering here the charge transport driven by voltages safely below the superconducting gap $e |V| < |\Delta|$ (practically we assume $|\Delta| \sim 100 \Gamma_L$). Nevertheless, even at such small voltage (of the order of a few $\Gamma_L$) the pile-up of electric charges in the electrodes and the dot affects the transport by shifting the chemical potentials and screening the charge on the dot. This is taken into account in the Hamiltonian (\ref{eq-Ham}) by the term $e U (\mathbf{r})$.

The effect has been considered first in mesoscopic normal systems by Altshuler and Khmelnitskii~\cite{altshuler1985}, B\"{u}ttiker with coworkers~\cite{buttiker1993,buttiker1997} and others~\cite{ma1998}. It has been also explored in the metal-superconductor (two-terminal) junctions~\cite{wang2001}. Here we follow [\onlinecite{wang2001}], assuming that the long range interactions modify the on-dot energy $\epsilon_0$ changing it to $\epsilon_0 - e U (\mathbf{r})$. In equilibrium the potential $U (\mathbf{r})$ has a constant value, which we denote by $U_{eq}$. In the presence of the applied voltages $V_\alpha$ (where $\alpha = L, R, S$) the deviations $\delta U = U (\mathbf{r}) - U_{eq}$, in the lowest order, would be a linear function
\begin{equation}
\delta U = \sum_\alpha \left( \frac{\partial U}{\partial V_\alpha} \right)_0 V_\alpha \; ,
\end{equation}
where $( ... )_0$ denotes the derivative with all voltages set to zero and the gauge invariance implies that $\sum_\alpha \left( \frac{\partial U}{\partial V_\alpha} \right) _0 = 1$~\cite{buttiker1993}. Our treatment here relies on the mean field like approximation. In the three terminal device with the quantum dot the single electron transport occurs between the left and right normal electrodes, while the (direct and crossed) Andreev processes involve the normal and superconducting electrodes. The currents (\ref{curr-gen-ET}, \ref{curr-gen-DAR}, \ref{curr-gen-CAR}) and the quantum dot charge $n = 2 \int \frac{dE}{2 \pi} \left[ |G_{11}^r|^2 ( \Gamma_L f_L + \Gamma_R f_R ) + | G_{12}^r |^2 ( \Gamma_L \tilde{f}_L + \Gamma_R \tilde{f}_R ) \right] $ depend on the screening potential $U (\mathbf{r})$. During the flow of carriers the deviations of $\delta U$ from the equilibrium value $U_{eq}$ can be related to the change of the charge carriers $\delta n$ by the capacitance equation
%
$\delta n = C \delta U$,
%
where $C$ is capacity of the system. The charge density as well as all currents depend on the voltages and $\delta U$. This allows to write the relation between $\delta n = n - n_{eq}$, where $n_{eq}$ denotes the equilibrium (i.e. calculated for all voltages set to zero) value of the charge
\begin{equation}
\delta n =\sum_\alpha \left( \frac{\partial n}{\partial V_\alpha} \right)_{0} V_\alpha - \Pi \; \delta U \; ,
\end{equation}
where $\Pi$ denotes the Lindhard function. Combining these equations we solve for $( \frac{\partial U}{\partial V_\alpha} ) _0$ known in the literature as the characteristic potentials and conveniently denoted by $u_\alpha$. They describe the response of the system to the applied voltages. One finds
\begin{equation}
u_\alpha = \frac{1}{C + \Pi} \left( \frac{\partial n}{\partial V_\alpha} \right)_0 \; .
\end{equation}

For the analysis of voltages induced in the $R$ electrode as a result of current flowing in the $L-S$ branch of the system we need both $u_L$ and $u_R$. As in the earlier work~[\onlinecite{wang2001}] we assume $C = 0$ in the following. The inspection of the formula for $n$ reveals that for the symmetric coupling $\Gamma_L = \Gamma_R$ the functions of both electrodes take on the same value $u_L = u_R$. The characteristic potentials enter the expression for the Green functions and as a result modify the relation shown in the figure~\ref{vr-vs-vl}. The modification is especially severe for $V_L > \Gamma_L$.

Let us note that $\Pi = - \left( \frac{\delta n}{\delta U} \right)_0$ is obtained from matrix elements $G_{11}^r$ and $G_{12}^r$ of the the Green functions as they depend on the potential $U$. The calculation of the characteristic potentials $u_{L / R}$ require the derivatives of $n$ with respect to voltages $V_{L / R}$, which enter the distribution functions. The characteristic functions define in turn the potential $U = u_L V_L + u_R V_R$, which has to be introduced into the Green functions entering the expressions (\ref{curr-gen-ET}, \ref{curr-gen-DAR}, \ref{curr-gen-CAR}) for the currents.

\section*{Acknowledgements}

Authors acknowledge M. Urbaniak for the technical assistance. This work is supported by the National Science Centre under the contracts DEC-2012/05/B/ST3/03208 (GM, BRB), DEC-2014/13/B/ST3/04451 (TD), DEC-2011/01/B/ST3/04428 (KIW).

\end{document}